\documentclass[10pt, journal, twocolumn, letterpaper]{IEEEtran}

\makeatletter

\newcommand{\Rmnum}[1]{\expandafter\@slowromancap\romannumeral #1@}
\makeatother
\usepackage{array}
\usepackage{url}
\usepackage{multirow}
\usepackage{diagbox}
\usepackage{amsthm,amsmath,amsfonts,syntonly}
\usepackage{amssymb}
\usepackage{epsfig} 
\usepackage{epstopdf}
\usepackage{epsf, graphics,graphicx}
\usepackage{cite, bm,color,url,textcomp}
\usepackage{float}
\usepackage{algorithm}
\usepackage{algorithmic}
\usepackage{lipsum}
\usepackage{mathrsfs}
\usepackage{arydshln}
\usepackage{stfloats}
\newcommand\blfootnote[1]{%
  \begingroup
  \renewcommand\thefootnote{}\footnote{#1}%
  \addtocounter{footnote}{-1}%
  \endgroup
}

\begin{document}
\newcommand{\fs}{\hspace{0.07in}}
\newcommand{\bs}{\hspace{-0.1in}}
\newcommand{\re}{{\rm Re} \, }
\newcommand{\e}{{\rm E} \, }
\newcommand{\p}{{\rm P} \, }
\newcommand{\cn}{{\cal CN} \, }
\newcommand{\n}{{\cal N} \, }
\newcommand{\ba}{\begin{array}}
\newcommand{\ea}{\end{array}}
\newcommand{\be}{\begin{displaymath}}
\newcommand{\ee}{\end{displaymath}}
\newcommand{\ben}{\begin{equation}}
\newcommand{\een}{\end{equation}}
\newcommand{\bena}{\begin{eqnarray}}
\newcommand{\eena}{\end{eqnarray}}
\newcommand{\beqa}{\begin{eqnarray*}}
\newcommand{\enqa}{\end{eqnarray*}}
\newcommand{\f}{\frac}
\newcommand{\bc}{\begin{center}}
\newcommand{\ec}{\end{center}}
\newcommand{\bi}{\begin{itemize}}
\newcommand{\ei}{\end{itemize}}
\newcommand{\benu}{\begin{enumerate}}
\newcommand{\eenu}{\end{enumerate}}
\newcommand{\bdes}{\begin{description}}
\newcommand{\edes}{\end{description}}
\newcommand{\bt}{\begin{tabular}}
\newcommand{\et}{\end{tabular}}
\newcommand{\vs}{\vspace}
\newcommand{\hs}{\hspace}
\newcommand{\sort}{\rm sort \,}

\newcommand \thetabf{{\mbox{\boldmath$\theta$\unboldmath}}}
\newcommand{\Phibf}{\mbox{${\bf \Phi}$}}
\newcommand{\Psibf}{\mbox{${\bf \Psi}$}}
\newcommand \alphabf{\mbox{\boldmath$\alpha$\unboldmath}}
\newcommand \betabf{\mbox{\boldmath$\beta$\unboldmath}}
\newcommand \gammabf{\mbox{\boldmath$\gamma$\unboldmath}}
\newcommand \deltabf{\mbox{\boldmath$\delta$\unboldmath}}
\newcommand \epsilonbf{\mbox{\boldmath$\epsilon$\unboldmath}}
\newcommand \zetabf{\mbox{\boldmath$\zeta$\unboldmath}}
\newcommand \etabf{\mbox{\boldmath$\eta$\unboldmath}}
\newcommand \iotabf{\mbox{\boldmath$\iota$\unboldmath}}
\newcommand \kappabf{\mbox{\boldmath$\kappa$\unboldmath}}
\newcommand \lambdabf{\mbox{\boldmath$\lambda$\unboldmath}}
\newcommand \mubf{\mbox{\boldmath$\mu$\unboldmath}}
\newcommand \nubf{\mbox{\boldmath$\nu$\unboldmath}}
\newcommand \xibf{\mbox{\boldmath$\xi$\unboldmath}}
\newcommand \pibf{\mbox{\boldmath$\pi$\unboldmath}}
\newcommand \rhobf{\mbox{\boldmath$\rho$\unboldmath}}
\newcommand \sigmabf{\mbox{\boldmath$\sigma$\unboldmath}}
\newcommand \taubf{\mbox{\boldmath$\tau$\unboldmath}}
\newcommand \upsilonbf{\mbox{\boldmath$\upsilon$\unboldmath}}
\newcommand \phibf{\mbox{\boldmath$\phi$\unboldmath}}
\newcommand \varphibf{\mbox{\boldmath$\varphi$\unboldmath}}
\newcommand \chibf{\mbox{\boldmath$\chi$\unboldmath}}
\newcommand \psibf{\mbox{\boldmath$\psi$\unboldmath}}
\newcommand \omegabf{\mbox{\boldmath$\omega$\unboldmath}}
\newcommand \Sigmabf{\hbox{$\bf \Sigma$}}
\newcommand \Upsilonbf{\hbox{$\bf \Upsilon$}}
\newcommand \Omegabf{\hbox{$\bf \Omega$}}
\newcommand \Deltabf{\hbox{$\bf \Delta$}}
\newcommand \Gammabf{\hbox{$\bf \Gamma$}}
\newcommand \Thetabf{\hbox{$\bf \Theta$}}
\newcommand \Lambdabf{\hbox{$\bf \Lambda$}}
\newcommand \Xibf{\hbox{\bf$\Xi$}}
\newcommand \Pibf{\hbox{\bm$\Pi$}}
\newcommand \abf{{\bf a}}
\newcommand \bbf{{\bf b}}
\newcommand \cbf{{\bf c}}
\newcommand \dbf{{\bf d}}
\newcommand \ebf{{\bf e}}
\newcommand \fbf{{\bf f}}
\newcommand \gbf{{\bf g}}
\newcommand \hbf{{\bf h}}
\newcommand \ibf{{\bf i}}
\newcommand \jbf{{\bf j}}
\newcommand \kbf{{\bf k}}
\newcommand \lbf{{\bf l}}
\newcommand \mbf{{\bf m}}
\newcommand \nbf{{\bf n}}
\newcommand \obf{{\bf o}}
\newcommand \pbf{{\bf p}}
\newcommand \qbf{{\bf q}}
\newcommand \rbf{{\bf r}}
\newcommand \sbf{{\bf s}}
\newcommand \tbf{{\bf t}}
\newcommand \ubf{{\bf u}}
\newcommand \vbf{{\bf v}}
\newcommand \wbf{{\bf w}}
\newcommand \xbf{{\bf x}}
\newcommand \ybf{{\bf y}}
\newcommand \zbf{{\bf z}}
\newcommand \rbfa{{\bf r}}
\newcommand \xbfa{{\bf x}}
\newcommand \ybfa{{\bf y}}
\newcommand \Abf{{\bf A}}
\newcommand \Bbf{{\bf B}}
\newcommand \Cbf{{\bf C}}
\newcommand \Dbf{{\bf D}}
\newcommand \Ebf{{\bf E}}
\newcommand \Fbf{{\bf F}}
\newcommand \Gbf{{\bf G}}
\newcommand \Hbf{{\bf H}}
\newcommand \Ibf{{\bf I}}
\newcommand \Jbf{{\bf J}}
\newcommand \Kbf{{\bf K}}
\newcommand \Lbf{{\bf L}}
\newcommand \Mbf{{\bf M}}
\newcommand \Nbf{{\bf N}}
\newcommand \Obf{{\bf O}}
\newcommand \Pbf{{\bf P}}
\newcommand \Qbf{{\bf Q}}
\newcommand \Rbf{{\bf R}}
\newcommand \Sbf{{\bf S}}
\newcommand \Tbf{{\bf T}}
\newcommand \Ubf{{\bf U}}
\newcommand \Vbf{{\bf V}}
\newcommand \Wbf{{\bf W}}
\newcommand \Xbf{{\bf X}}
\newcommand \Ybf{{\bf Y}}
\newcommand \Zbf{{\bf Z}}
\newcommand \Omegabbf{{\bf \Omega}}
\newcommand \Rssbf{{\bf R_{ss}}}
\newcommand \Ryybf{{\bf R_{yy}}}
\newcommand \Cset{{\cal C}}
\newcommand \Rset{{\cal R}}
\newcommand \Zset{{\cal Z}}
\newcommand \Sset{{\cal S}}

\newcommand \Dset{{\cal D}}
\newcommand \Fset{{\cal F}}
\newcommand \Wset{{\cal W}}
\newcommand \Qset{{\cal Q}}
\newcommand \Tset{{\cal T}}

\newcommand{\otheta}{\stackrel{\circ}{\theta}}
\newcommand{\defeq}{\stackrel{\bigtriangleup}{=}}
\newcommand{\oabf}{{\bf \breve{a}}}
\newcommand{\odbf}{{\bf \breve{d}}}
\newcommand{\oDbf}{{\bf \breve{D}}}
\newcommand{\oAbf}{{\bf \breve{A}}}
\renewcommand \vec{{\mbox{vec}}}
\newcommand{\Acalbf}{\bf {\cal A}}
\newcommand{\calZbf}{\mbox{\boldmath $\cal Z$}}
\newcommand{\feop}{\hfill \rule{2mm}{2mm} \\}
\newtheorem{theorem}{Theorem}[section]

\newcommand{\Rnum}{{\mathbb R}}
\newcommand{\Cnum}{{\mathbb C}}
\newcommand{\Znum}{{\mathbb Z}}
\newcommand{\Enum}{{\mathbb E}}
\newcommand{\Nnum}{{\mathbb N}}

\newcommand{\Ical}{{\cal I}}
\newcommand{\Mcal}{{\cal M}}
\newcommand{\Pcal}{{\cal P}}
\newcommand{\Ccal}{{\cal C}}
\newcommand{\Dcal}{{\cal D}}
\newcommand{\Hcal}{{\cal H}}
\newcommand{\Ocal}{{\cal O}}
\newcommand{\Rcal}{{\cal R}}
\newcommand{\Zcal}{{\cal Z}}
\newcommand{\Xcal}{{\cal X}}
\newcommand{\Qcal}{{\cal Q}}
\newcommand{\Gcal}{{\cal G}}
\newcommand{\zzbf}{{\bf 0}}
\newcommand{\zebf}{{\bf 0}}

\newcommand{\eop}{\hfill $\Box$}

\newcommand{\gss}{\mathop{}\limits}
\newcommand{\gs}{\mathop{\gss_<^>}\limits}

\newcommand{\circlambda}{\mbox{$\Lambda$
             \kern-.85em\raise1.5ex
             \hbox{$\scriptstyle{\circ}$}}\,}

\newcommand{\tr}{\mathop{\rm tr}}
\newcommand{\var}{\mathop{\rm var}}
\newcommand{\cov}{\mathop{\rm cov}}
\newcommand{\diag}{\mathop{\rm diag}}
\newcommand{\blkdiag}{\mathop{\rm blkdiag}}
\def\rank{\mathop{\rm rank}\nolimits}
\newcommand{\ra}{\rightarrow}
\newcommand{\ul}{\underline}
\def\Pr{\mathop{\rm Pr}}
\def\Re{\mathop{\rm Re}}
\def\Im{\mathop{\rm Im}}

\def\submbox#1{_{\mbox{\footnotesize #1}}}
\def\supmbox#1{^{\mbox{\footnotesize #1}}}

%
\newtheorem{Theorem}{Theorem}[section]
\newtheorem{Definition}[Theorem]{Definition}
\newtheorem{Proposition}[Theorem]{Proposition}
\newtheorem{Lemma}[Theorem]{Lemma}
\newtheorem{Corollary}[Theorem]{Corollary}
%
%
\newcommand{\ThmRef}[1]{\ref{thm:#1}}
\newcommand{\ThmLabel}[1]{\label{thm:#1}}
\newcommand{\DefRef}[1]{\ref{def:#1}}
\newcommand{\DefLabel}[1]{\label{def:#1}}
\newcommand{\PropRef}[1]{\ref{prop:#1}}
\newcommand{\PropLabel}[1]{\label{prop:#1}}
\newcommand{\LemRef}[1]{\ref{lem:#1}}
\newcommand{\LemLabel}[1]{\label{lem:#1}}
%

\newcommand \bbs{{\boldsymbol b}}
\newcommand \cbs{{\boldsymbol c}}
\newcommand \dbs{{\boldsymbol d}}
\newcommand \ebs{{\boldsymbol e}}
\newcommand \fbs{{\boldsymbol f}}
\newcommand \gbs{{\boldsymbol g}}
\newcommand \hbs{{\boldsymbol h}}
\newcommand \ibs{{\boldsymbol i}}
\newcommand \jbs{{\boldsymbol j}}
\newcommand \kbs{{\boldsymbol k}}
\newcommand \lbs{{\boldsymbol l}}
\newcommand \mbs{{\boldsymbol m}}
\newcommand \nbs{{\boldsymbol n}}
\newcommand \obs{{\boldsymbol o}}
\newcommand \pbs{{\boldsymbol p}}
\newcommand \qbs{{\boldsymbol q}}
\newcommand \rbs{{\boldsymbol r}}
\newcommand \sbs{{\boldsymbol s}}
\newcommand \tbs{{\boldsymbol t}}
\newcommand \ubs{{\boldsymbol u}}
\newcommand \vbs{{\boldsymbol v}}
\newcommand \wbs{{\boldsymbol w}}
\newcommand \xbs{{\boldsymbol x}}
\newcommand \ybs{{\boldsymbol y}}
\newcommand \zbs{{\boldsymbol z}}

\newcommand \Bbs{{\boldsymbol B}}
\newcommand \Cbs{{\boldsymbol C}}
\newcommand \Dbs{{\boldsymbol D}}
\newcommand \Ebs{{\boldsymbol E}}
\newcommand \Fbs{{\boldsymbol F}}
\newcommand \Gbs{{\boldsymbol G}}
\newcommand \Hbs{{\boldsymbol H}}
\newcommand \Ibs{{\boldsymbol I}}
\newcommand \Jbs{{\boldsymbol J}}
\newcommand \Kbs{{\boldsymbol K}}
\newcommand \Lbs{{\boldsymbol L}}
\newcommand \Mbs{{\boldsymbol M}}
\newcommand \Nbs{{\boldsymbol N}}
\newcommand \Obs{{\boldsymbol O}}
\newcommand \Pbs{{\boldsymbol P}}
\newcommand \Qbs{{\boldsymbol Q}}
\newcommand \Rbs{{\boldsymbol R}}
\newcommand \Sbs{{\boldsymbol S}}
\newcommand \Tbs{{\boldsymbol T}}
\newcommand \Ubs{{\boldsymbol U}}
\newcommand \Vbs{{\boldsymbol V}}
\newcommand \Wbs{{\boldsymbol W}}
\newcommand \Xbs{{\boldsymbol X}}
\newcommand \Ybs{{\boldsymbol Y}}
\newcommand \Zbs{{\boldsymbol Z}}

\newcommand \Absolute[1]{\left\lvert #1 \right\rvert}

\title{Over-The-Air Phase Calibration of Spaceborne Phased Array for LEO Satellite Communications}
\author{Wei Zhang, Ding Chen, Bin Zhou
}

\maketitle
\blfootnote{
Wei Zhang, Ding Chen and Bin Zhou are with the Shanghai Institute of Microsystem and Information Technology and the Science and Technology on Micro-System Laboratory, Chinese Academy of Sciences, Shanghai (emails: wzhang@mail.sim.ac.cn, ding.chen@mail.sim.ac.cn, bin.zhou@mail.sim.ac.cn). ({\em Corresponding author: Bin Zhou}).
}

\begin{abstract}
To avoid the unpredictable phase deviations of the spaceborne phased array (SPA), this paper considers the over-the-air (OTA) phase calibration of the SPA for the low earth orbit (LEO) satellite communications, where the phase deviations of the SPA and the unknown channel are jointly estimated with multiple transmissions of the pilots. Moreover, the Cram\'{e}r Rao Bound (CRB) is derived, and the optimization of beam patterns is also presented to lower the root mean squared error (RMSE) of the OTA calibration. The simulation results verify the effectiveness of the proposed OTA phase calibration algorithm as the RMSEs of the phase estimates closely approach the corresponding CRB, and the beam pattern optimization scheme is also validated for more than $4$dB gain of SNR over the randomly generated beam patterns.
\end{abstract}
\begin{IEEEkeywords}
spaceborne phased array (SPA), low earth orbit (LEO) satellite communications, over-the-air (OTA), Cram\'{e}r Rao Bound (CRB).
\end{IEEEkeywords}

%
\IEEEpeerreviewmaketitle

\section{Introduction}
Low earth orbit (LEO) satellite communications are undergoing a revolutionary transformation, driven by satellite constellations (e.g., Starlink) delivering global coverage \cite{9800119}. As a core equipment for the LEO  communications, spaceborne phased arrays (SPA) has millisecond beam agility and multi-beam spatial multiplexing capability \cite{9955558}. In practice, however, the SPA are unavoidably subject to unpredictable phase deviations owing to aging and cumulative radiation damage in space \cite{10797644}. Although robust algorithm for  these hardware impairments can be developed \cite{10844052}\cite{10535263}, the data rate will still be reduced compared with the condition without any phase deviations. Thus these phase deviations need to be calibrated periodically.

Conventional phased array calibration has been widely investigated for decades. From the aspect of the number of array elements measured per time, the existing calibration methods can be divided into the methods that depend on measurements of a single element \cite{4410650508}\cite{923310} and the methods based on simultaneous measurements of an array of elements \cite{4558321}\cite{5979144}; from the aspect of signal characteristics, the existing methods can also be divided into power variation-based methods \cite{4558321}\cite{10366806} and complex value-based methods \cite{8693562}\cite{9423585}. The aforementioned methods are conducted in the anechoic chamber which is almost a noise-free environment, and thus these techniques cannot directly calibrate the phased array for communication systems exposed to complex electromagnetic environments, especially in space environment. The conventional in-orbit calibration approach is to maintain a table that records the deviations measured at different temperatures at ground-based anechoic chamber, and the deviations are calibrated in-orbit through looking up the table according to the practical temperature in space, which only applies to the phase deviations caused by temperature drift.

As for device aging and the cumulative radiation damage, the over-the-air (OTA) phase calibration of the SPA is necessary. The paper \cite{9248044} utilized the multisine signal to calibrate the antenna array for the MIMO system, which only applies to transmitting (Tx) antenna arrays and cannot be extended to the OTA calibration of large scale phased arrays. The paper \cite{9056550} first considered the OTA phase calibration for mmWave hybrid MIMO systems but without providing a scheme on how to select the beam patterns. The paper \cite{9810495} proposed an OTA phase shifter network calibration method for hybrid multi-user systems where the optimal beam patterns are designed only for calibrating two-element phased array. To the best of our knowledge, the beamforming techniques for communications have not been considered in the in-orbit OTA calibration of the large-scale SPA before.

This paper considers the in-orbit OTA phase calibration of the hybrid analog-digital SPA for the LEO satellite communication systems. We propose a block coordinate descent algorithm to jointly estimate the phase deviations and the channel state information (CSI) based on the measured effective channels between the satellite and terrestrial terminal, and also optimize the beam patterns for more accurate calibration. The simulation results verify the proposed calibration algorithm by showing that the root mean squared error (RMSE) closely approaches the Cram\'{e}r Rao Bound (CRB) and the optimized beam patterns can offer more than $4$dB gain of SNR than that of the random beam patterns.

\section{Signal Model and Problem Formulation} \label{SEC2}
\subsection{Signal Model}
\begin{figure}[htb]
\centering
{\psfig{figure=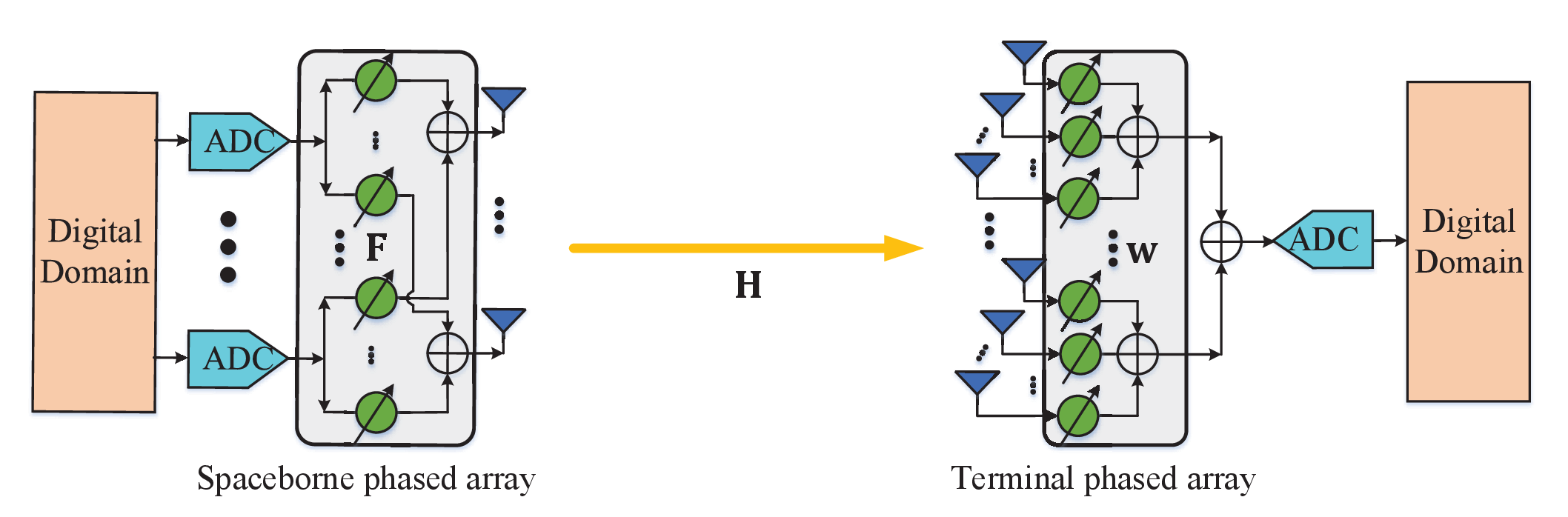,width= 3.5in}}
 \caption{Signal transmission model of phase calibration for the SPA.}
\label{fig.sigModel}
\end{figure}
Consider a low earth orbit (LEO)  satellite communication system comprised of an $M_t$-antenna spaceborne transmitter and an $M_r$-antenna terminal, both of which are equipped with a uniform planar antenna array as shown in Fig. \ref{fig.sigModel}. For the LEO satellite transmitter, its spaceborne phased array (SPA) of hybrid analog-digital architecture can be denoted as $\Fbf = \left[\fbf_1,\fbf_2,\dots,\fbf_{N_{\rm RF}}\right] \in {\Cnum}^{M_t\times N_{\rm RF}}$, where $\fbf_{n} \in {\Sset}^{M_t\times 1}, n = 1,2,\dots, N_{\rm RF}$ with ${\Sset} = \{e^{j\zeta}|\zeta\in[0,2\pi]\}$, $N_{\rm RF}$ represents the number of RF chains; For the terminal, the analog phased array is expressed as $\wbf\in {\Sset}^{M_r\times 1}$.

As the transmitter sends pilot $\Sbf \in {\mathbb C}^{N_{\rm RF}\times L}$ where $L$ is the pilot length and $\Sbf\Sbf^H = L\Ibf$, the receiver digital domain receives
\ben
\ybf = \sqrt{\beta}\wbf^H\Hbf\Fbf\Sbf + \zbf,
\label{equ.Y}
\een
where $\Hbf\in{\mathbb C}^{M_r\times M_{t}}$ represents the channel from the transmitter to the receiver and $\beta$ represents the pathloss between the satellite and the ground terminal;
$\Zbf\in{\mathbb C}^{N_{\rm RF}\times L}$ are complex Gaussian noise whose entries follow ${\cal CN}(0,\sigma^2)$. $\abf_{\rm UPA}(\theta, \phi)$ is the array response of the $x\times y$ uniform planar array (UPA) concerning the two-dimensional angle $(\theta, \phi)$: $\abf_{\rm UPA}(\theta,\phi) = \bar{\abf}(\theta,\phi)\otimes \check{\abf}(\phi)$, where $\bar{\abf}(\theta,\phi) = [1,e^{j\pi{\rm sin}(\theta){\rm sin}(\phi)},\dots, e^{j\pi(x-1){\rm sin}(\theta){\rm sin}(\phi)}]^T$ and $\check{\abf}(\phi) = [1,e^{j\pi{\rm cos}(\phi)},\dots,e^{j\pi(y-1){\rm cos}(\phi)}]^T$. The SPA has the array response as $\abf^H_{\rm UPA}(\theta_t,\phi_t)$ with $x = N_x$ and $y = N_y = \frac{M_t}{N_x}$; for the terminal phased array (TPA), we have its array response as $\abf_{\rm UPA}(\theta_r,\phi_r)$ with $x = M_x$ and $y = M_y = \frac{M_r}{M_x}$. Thus, we consider the channel between the LEO satellite and the terminal to be
\ben
\Hbf = \gamma\abf_{\rm UPA}(\theta_r,\phi_r)\abf^H_{\rm UPA}(\theta_t,\phi_t),
\label{equ.H}
\een
where $\gamma \sim \Ccal N(0,1)$ is the complex channel gain.

\subsection{Problem Formulation}
Owing to the extreme thermal cycling and cumulative radiation damage in space, the phases of the SPA may deviate from its nominal phase set. Considering the effect of the phase deviations, we can reformulate \eqref{equ.Y} as
\ben
\ybf = \sqrt{\beta}\wbf^H\Hbf(\Fbf\odot\Omegabf)\Sbf + \zbf,
\label{eq.y}
\een
where $\Omegabf = \left[\omegabf_1,\omegabf_2,\dots,\omegabf_{N_{\rm RF}}
\right]$ represents the phase shifts deviations with $\omegabf_{n}\in {\Sset}^{M_t\times 1}, n = 1,2,\dots, N_{\rm RF}$ and $\odot$ denotes the Hadamard product.
In the rest of this paper, we will jointly estimate $\Omegabf$ and the unknown channel $\Hbf$ based on \eqref{eq.y}, and then send the estimated $\Omegabf$ back to the satellite to calibrate its SPA.

Assume the LEO satellite sends pilots $\Sbf$ periodically for $K$ times, where the channel $\Hbf$ remains static, i.e., the pilots are transmitted within the coherent time-frequency block. Thus according to \eqref{eq.y}, the $k$-th transmission can be denoted as
\ben
\ybf_k =\sqrt{\beta}\wbf_k^H\Hbf(\Fbf_k\odot\Omegabf)\Sbf + \zbf_k,
\label{equ.Yk}
\een
where $\Fbf_k$ represents the $k$-th SPA nominal beamformer and $\wbf_k$ is the $k$-th TPA beamformer; $\ybf_k$ and $\zbf_k$ are the received signal and the noise in the $k$-th transmission.
Right-multiplying $\frac{1}{L}\Sbf^H$ on the both side of \eqref{equ.Yk}, we have
\ben
\tilde{\ybf}_k = \sqrt{\beta}\wbf_k^H\Hbf(\Fbf_k\odot \Omegabf) + \tilde{\zbf}_k,
\label{equ.Ysk}
\een
where $\tilde{\ybf}_k = \frac{1}{L}{\ybf}_k\Sbf^H \in{\Cnum}^{1 \times N_{\rm RF}}$; $\tilde{\zbf}_k = \frac{1}{L}{\zbf}_k\Sbf^H\in{\Cnum}^{1 \times N_{\rm RF}}$, of which the entries follow the distribution of ${\cal CN}(0,\frac{\sigma^2}{L})$. Hence the signal to noise ratio (SNR) can be denoted as $\frac{L}{\sigma^2}$.

Hence we can formulate an optimization problem regarding both $\Omegabf$ and $\Hbf$, i.e.,
\begin{align}
\mathop{\min}_{\Omegabf,\Hbf} \ &\sum_{k=1}^K \|\tilde{\ybf}_k -\sqrt{\beta}\wbf_k^H\Hbf(\Fbf_k\odot \Omegabf)\|_2^2.
\label{equ.objFuncOrid}
\end{align}
The problem \eqref{equ.objFuncOrid} is a non-convex due to the unit-phase shifts, to which the optimal solution is difficult to obtain. This paper proposes a block coordinate descent algorithm to solve problem \eqref{equ.objFuncOrid}, as detailed in the following sections.

\section{Over-The-Air Phases Calibration Algorithm}
In this section, we propose a block coordinate descent algorithm to solve \eqref{equ.objFuncOrid}: we first initialize $\Omegabf$ with all entries being $1$, and estimate the channel $\Hbf$, then given the estimated $\Hbf$ and update $\Omegabf$; as the above iteration goes on, the final result can be obtained until \eqref{equ.objFuncOrid} converges. Given the estimated $\Omegabf$ and $\Hbf$, we also point out the phase ambiguity between them, and remove it by pre-assigning values to the corresponding variables.

\subsection{Channel Estimation}\label{sec.CE}
Given that $\Omegabf$ is fixed, we can denote
\ben
\Abf_k = \Fbf_k\odot\Omegabf.
\label{eq.Ak}
\een
Thus inserting \eqref{equ.H} and \eqref{eq.Ak} into \eqref{equ.Ysk}, we can further obtain
\ben
\tilde{\ybf}_k =\sqrt{\beta}\gamma\wbf_k^H\abf_{\rm UPA}(\theta_r,\phi_r)\abf^H_{\rm UPA}(\theta_t,\phi_t)\Abf_k + \tilde{\zbf}_k.
\label{eq.tykl}
\een
Applying the formula ${\rm vec}(\Xbf\Ybf\Zbf) = (\Zbf^T\otimes\Xbf){\rm vec}(\Ybf)$ to \eqref{eq.tykl} and denoting $\Bbf_k = \sqrt{\beta}(\Abf_k^T\otimes\wbf_k^H)\in \Cnum^{N_{\rm RF}\times M_rM_t}$ where $\otimes$ represents Kronecker product and ${\rm vec}(\cdot)$ transforms a matrix into a column vector by concatenating its columns, we have
\ben
\tilde{\ybf}_{k, \rm vec} = \gamma\Bbf_k\abf_{{\rm vec}}(\theta_r,\phi_r,\theta_t,\phi_t) + \tilde{\zbf}_{k, \rm vec},
\label{equ.tyvec}
\een
where $\abf_{{\rm vec}}(\theta_r,\phi_r,\theta_t,\phi_t) = {\rm vec}\left(\abf_{\rm UPA}(\theta_r,\phi_r)\abf^H_{\rm UPA}(\theta_t,\phi_t)\right)$, $\tilde{\ybf}_{k,{\rm vec}} = {\rm vec}(\tilde{\ybf}_k)$, and $\tilde{\zbf}_{k, \rm vec} = {\rm vec}(\tilde{\zbf}_k)$. We rewrite
$\abf_{{\rm vec}}(\theta_r,\phi_r,\theta_t,\phi_t)$ as $\abf_{{\rm vec}}$ for short in the rest of this paper.

According to \eqref{equ.tyvec}, denoting $\Bbf = [\Bbf_1^T,\Bbf_2^T,\dots,\Bbf_K^T]^T$, we can have the equivalent form of \eqref{equ.objFuncOrid} as
\begin{align}
\mathop{\min}_{\gamma, \theta_r, \phi_r, \theta_t, \phi_t} \ &\|\tilde{\ybf}_{\rm vec} - \gamma\Bbf\abf_{{\rm vec}}\|_2^2,
\label{equ.objFuncGv2}
\end{align}
where $\tilde{\ybf}_{\rm vec} = [\tilde{\ybf}_{1, \rm vec}^T,\tilde{\ybf}_{2, \rm vec}^T,\dots,\tilde{\ybf}_{K, \rm vec}^T]^T$. We can obtain the least-squared (LS) solution to \eqref{equ.objFuncGv2} as $\gamma = \frac{\abf^H_{\rm vec}\Bbf^H\tilde{\ybf}_{\rm vec}}{\abf_{\rm vec}^H\Bbf^H\Bbf\abf_{\rm vec}}$. Inserting $\gamma$ back into \eqref{equ.objFuncGv2} yields that
\begin{align}
\mathop{\max}_{\theta_r,\phi_r,\theta_t,\phi_t} \ & f(\theta_r,\phi_r,\theta_t,\phi_t) \triangleq \frac{\|\tilde{\ybf}_{\rm vec}^H\Bbf\abf_{\rm vec}\|_2^2}{\|\Bbf\abf_{\rm vec}\|_2^2}.
\label{equ.objFuncGv3}
\end{align}
\subsubsection{Coarse Solution to \eqref{equ.objFuncGv3} Using 4D-FFTs} \label{sec.4Dfft}
we reformulate \eqref{equ.objFuncGv3} into
\begin{align}
\mathop{\max}_{\theta_r,\phi_r,\theta_t,\phi_t} \ &  f(\theta_r,\phi_r,\theta_t,\phi_t) \triangleq \frac{\|\tilde{\ybf}_{\rm vec}^H\Bbf\abf_{\rm vec}\|_2^2}{\sum_{m=1}^{M_rM_t}\|\bbf_m^T\abf_{\rm vec}\|_2^2},
\label{equ.objFuncGv4}
\end{align}
where $\bbf_m^T$ represents the $m$-th row of $\Bbf$. For the denominator of \eqref{equ.objFuncGv3},  $\|\bbf_m^T\abf_{\rm vec}\|_2^2$ can be calculated as shown in \eqref{equ.qmavec} where $\bbf_{m,n_x,n_y,m_x,m_y}$ represents the $\left(((n_x-1)N_y +  n_y-1)M_r + (m_x-1)M_y + m_y\right)$-th elements of $\bbf_m$. According to \eqref{equ.qmavec}, we can evaluate the value of $\|\bbf_m^T\abf_{\rm vec}\|_2^2$ over the mesh grids with $N_{\theta_r}\times N_{\phi_r}\times N_{\theta_t}\times N_{\phi_t}$-point 4D-FFT, yielding a four-way tensor $\Qset_m\in\Rnum^{N_{\theta_r}\times N_{\phi_r}\times N_{\theta_t}\times N_{\phi_t}}$. For the numerator of \eqref{equ.objFuncGv3}, we can also apply 4D-FFT to  $\|\tilde{\ybf}_{\rm vec}^H\Bbf\abf_{\rm vec}\|_2^2$ and obtain a four-way tensor $\Rset\in\Rnum^{N_{\theta_r}\times N_{\phi_r}\times N_{\theta_t}\times N_{\phi_t}}$. Then letting $\Qset = \sum_{m=1}^{M_rM_t}\Qset_m$ and dividing $\Qset$ by $\Rset$ in an element-wise manner, we can obtain $\Tset\in\Rnum^{N_{\theta_r}\times N_{\phi_r}\times N_{\theta_t}\times N_{\phi_t}}$, of which the largest entry located at $(n_{\theta_r},n_{\phi_r},n_{\theta_t},n_{\phi_t})$ represents the maximum of \eqref{equ.objFuncGv4}, leading the coarse frequency points as $f_{\theta_r} = \frac{n_{\theta_r}-1}{N_{\theta_r}},\ f_{\phi_r} = \frac{n_{\phi_r}-1}{N_{\phi_r}},
f_{\theta_t} = \frac{n_{\theta_t}-1}{N_{\theta_t}},\ f_{\phi_t} = \frac{n_{\phi_t}-1}{N_{\phi_t}}.$
But according to \eqref{equ.qmavec}, the real frequencies are $\frac{1}{2}{\rm cos}(\phi_t)$, $\frac{1}{2}{\rm sin}(\theta_t){\rm sin}(\phi_t)$,$-\frac{1}{2}{\rm cos}(\phi_r)$, and $-\frac{1}{2}{\rm sin}(\theta_r){\rm sin}(\phi_r)$, all ranging from $-1/2$ to $1/2$. Hence we need to adjust the frequencies as
$f = \left\{
\ba{ll}
f, & f < \frac{1}{2}; \\
f-1,& f \ge \frac{1}{2},
\ea
\right.$
and then $\theta_r$, $\phi_r$, $\theta_t$, and $\phi_t$ can be estimated as
$\phi_r = {\rm cos}^{-1}(-2f_{\phi_r}), \theta_r = {\rm sin}^{-1}(-\frac{2f_{\theta_r}}{{\rm sin}(\phi_r)}),\phi_t = {\rm cos}^{-1}(2f_{\phi_t}), \theta_t = {\rm sin}^{-1}(\frac{2f_{\theta_t}}{{\rm sin}(\phi_t)}).$

\begin{figure*}[tb]
\ben
\begin{split}
\bbf_m^T\abf_{\rm vec} &= \sum_{n_y=1}^{N_y}\left(\sum_{n_x=1}^{N_x}\tilde{b}_{m,n_x,n_y}e^{j\pi(n-1){\rm sin(\theta_r)sin(\phi_r)}}\right)e^{j\pi(n_y-1){\rm cos}(\phi_r)}, \\
\tilde{b}_{m,n_x,n_y} &= \sum_{m_y=1}^{M_y}\left(\sum_{m_x=1}^{M_x}\bbf_{m,n_x,n_y,m_x,m_y}e^{-j\pi(m_x-1){\rm sin(\theta_t)sin(\phi_t)}}\right)e^{-j\pi(m_y-1){\rm cos}(\phi_t)}.
\end{split}
\label{equ.qmavec}
\een
\end{figure*}

\subsubsection{Refined Solution to \eqref{equ.objFuncGv3} Using Backtracking Line Search}
given the coarse solution to \eqref{equ.objFuncGv3}, we can find a more accurate solution to \eqref{equ.objFuncGv3} using the backtracking line search method. The search process in the $i$-th iteration is
$\zetabf_{i+1} = \zetabf_{i} - \alpha_i\nabla f(\zetabf)$,
where $\zetabf = [\theta_r,\phi_r,\theta_t,\phi_t]^T$, $\nabla f(\zetabf) = \left[
\frac{\partial f}{\partial \theta_r},
\frac{\partial f}{\partial \phi_r},
\frac{\partial f}{\partial \theta_t},
\frac{\partial f}{\partial \phi_t}
\right]^T$ where $\alpha_i$ can be determined by the Armijo-Goldstein condition.

\subsection{Phases Estimation}\label{sec.PE}
Given an estimated $\Hbf$ and denoting $\abf_{k,n}$ as the $n$-th column of $\Abf_k$ respectively, we have from \eqref{eq.tykl} that
\ben
\tilde{y}_{k,n} = \sqrt{\beta}\wbf_k^H\Hbf\abf_{k,n} + \tilde{z}_{k,n},
\label{equ.tyknx}
\een
where $\tilde{y}_{k,n}$ and $\tilde{z}_{k,n}$ are the $n$-th element of $\tilde{\ybf}_k$ and $\tilde{\zbf}_k$. With $\fbf_{k,n}$ being the $n$-th column of $\Fbf_k$, we have from \eqref{eq.Ak} that
\ben
\abf_{k,n} = \fbf_{k,n}\odot \omegabf_{n} = \diag(\fbf_{k,n})\omegabf_{n},
\label{eq.aknx}
\een
where $\diag(\fbf_{k,n})$ is a diagonal matrix with diagonal elements being the entries of $\fbf_{k,n}$. Inserting \eqref{eq.aknx} into \eqref{equ.tyknx} yields
\ben
\tilde{y}_{k,n} = \cbf^H_{k,n}\omegabf_{n} + \tilde{z}_{k,n}.
\label{equ.tyknxv2}
\een
where $\cbf_{k,n} = \sqrt{\beta}(\wbf_k^H\Hbf\diag(\fbf_{k,n}))^H$. Stacking $\cbf_{k,n}$'s as $\Cbf_{n} = [\cbf_{1,n},\cbf_{2,n},\dots,\cbf_{K,n}]^H\in \Cnum^{K\times M_t}$, we can obtain from \eqref{equ.tyknxv2} as
\ben
\bar{\ybf}_{n} = \Cbf_{n}\omegabf_{n} + \bar{\zbf}_{n},
\label{equ.barynx}
\een
where $\bar{\ybf}_{n} = [\tilde{y}_{1,n},\tilde{y}_{2,n},\dots,\tilde{y}_{K,n}]^T$ and
$\bar{\zbf}_{n} = [\tilde{z}_{1,n},\tilde{z}_{2,n},\dots,\tilde{z}_{K,n}]^T$. Hence  we need to solve
\begin{align}
\mathop{\max}_{\omegabf_{n}}  g(\omegabf_{n}) \triangleq \| \bar{\ybf}_{n} - \Cbf_{n}\omegabf_{n} \|_2^2, n
= 1,2,\dots, N_{\rm RF},
\label{equ.objFuncGv5}
\end{align}
from which we can obtain the Euclidean gradient as
\ben
\nabla g(\omegabf_{n}) = -\Cbf_{n}^H(\bar{\ybf}_{n} - \Cbf_{n}\omegabf_{n}).
\label{equ.nablag}
\een
As the unit-phase shifts of ${\omegabf}_{n},n=1,2,\dots,N_{\rm RF}$ defines a manifold
$\Mcal_{cc}^{M_t} \triangleq \left\{\xbf\in{\Cnum}^{M_t}: |\xbf(1)| =  \cdots = |\xbf({M_t})| = 1\right\}$ \cite{boumal2023intromanifolds}, we can utilize the Riemannian conjugate gradient (RCG) algorithm to obtain a sub-optimal solution to \eqref{equ.objFuncGv5} based on the Euclidean gradient from \eqref{equ.nablag}.

\subsection{Phase Ambiguity}\label{sec.pa}
According to \eqref{equ.objFuncOrid}, we can see that $\Omegabf, \Hbf$ are not uniquely determined as
\ben
\wbf_k^H\Hbf\left(\Fbf_k\odot\Omegabf\right) = \wbf_k^H e^{j\beta} \Hbf\Tbf\left(\Fbf_k\odot e^{-j\beta}(\Tbf^H\Omegabf)\right),
\een
where ${\Tbf} = \diag({\tbf})$, ${\tbf}=\bar{\abf}({\chi}_1,{\chi}_2)\otimes\check{\abf}({\chi}_2)$;
$\beta$, ${\chi}_1$, and ${\chi}_2$ are arbitrary phases, introducing ambiguity into the results. To remove this ambiguity, we assume that
$\beta = \angle{\Omegabf}_{1,1}, \
{\chi}_1 = \theta_{t}, \ {\chi}_2 = \phi_{t},$ where $\angle{\Omegabf}_{1,1}$ denotes the phase of $(1,1)$-th element of $\Omegabf$.

\subsection{Complexity Analysis}
We summarize the OTA calibration in Algorithm \ref{Algo.1}, and present an analysis on the computational complexity of the proposed OTA phase calibration method.
First, let us focus on the channel estimation in Sectoion \ref{sec.CE}: the computational complexity of $\Bbf$ is $T_B = \Ocal(KN_{\rm RF}M_rM_t)$; the computational complexity of initial coarse estimation mainly lies in the 4D-FFT, which is $T_I=\Ocal(M_rM_tN_{\rm FFT}{\rm log}_2(N_{\rm FFT}) + N_{\rm FFT}^4)$ given that $N_{\theta_r}= N_{\phi_r}= N_{\theta_t}= N_{\phi_t} = N_{\rm FFT}$; the complexity originates from the repeated calculation of $f(\zetabf)$ and its derivative $\nabla f(\zetabf)$, which can be approximated as $T_G = \Ocal(I_bKN_{\rm RF}M_tM_t)$ where $I_b$ represents the iteration times of the backtracking line search, thus the computational complexity of channel estimation for each iteration is about $T_C = T_B+T_I+T_G$. Second, for the phase estimation in Section \ref{sec.PE}, the complexity is $T_P = \Ocal(I_rKN_{\rm RF}M_t^2)$ where $I_r$ denotes the iteration times of the RCG algorithm.
Hence, the overall computational complexity can be expressed as $T = \Ocal\left(I_o\left( T_C + T_P\right)\right)$, where $I_o$ is the iteration times between the estimation of $\Omegabf$ and $\Hbf$. Under the condition that $M_t \approx M_r = K \gg I_o, I_r, I_b, N_{\rm RF}$, the computational complexity can be approximated as $T \approx \Ocal(M_t^3)$.
\begin{algorithm}[htb]
\caption{Algorithm for the OTA Calibration of the SPA}
\label{Algo.1}
\begin{algorithmic}[1]
\REQUIRE $\wbf_k$, $\Fbf_k,k=1,2,\dots,K$, $\tilde{\ybf}_{\rm vec}$, and $\bar{\ybf}_{n}$;
\ENSURE The phase deviations matrix $\Omegabf$ and the channel $\Hbf$;
\STATE Initialize $\Omegabf$ with matrix with all elements being $1$;
\STATE  Construct $\Bbf$ from $\Omegabf$, $\Fbf_k,k=1,2,\dots,K$, and $\wbf$.
\STATE Obtain $\Hbf$ using the methods proposed in Section \ref{sec.CE};
\WHILE {the cost function \eqref{equ.objFuncOrid} decreases by less than $1\%$}
\FOR {$n=1:N_{\rm RF}$}
\STATE Construct $\Dbf_{n}$ from $\wbf$, $\Hbf$, and $\Fbf_k,k=1,2,\dots, K$;
\STATE Obtain $\omegabf_{n}$ by solving \eqref{equ.objFuncGv5};
\ENDFOR
\STATE Fix $\Omegabf$ and refine $\Hbf$ by solving \eqref{equ.objFuncGv3};
\ENDWHILE
\STATE Remove the ambiguity between $\Omegabf$ and $\Hbf$ according to Section \ref{sec.pa};
\end{algorithmic}
\end{algorithm}

\section{Cram\'{e}r Rao Bounds}
This section will provide the Cram\'{e}r Rao Bound (CRB) for the phase calibration algorithm, which can act as a benchmark to gauge the performance of the proposed calibration algorithm. We also propose an algorithm to optimize the beam pattern $\fbf_{k,n},k=1,2,\dots,K, n = 1,2,\dots,N_{\rm RF}$ to achieve better phase calibration performance.

\subsection{Cram\'{e}r Rao Bounds}
Considering the phase ambiguity, we collect the rest phases of elements of $\Omegabf$ into the vector $\pbf$, where $\pbf = \left[\angle{\Omegabf_{2,1}},\dots,\angle{\Omegabf_{M_t,1}},\dots, \angle{\Omegabf_{1,2}},\dots,\angle{\Omegabf_{M_t-1,2}},\angle{\Omegabf_{M_t,N_{\rm RF}}},\right.$
$\left. \dots, \angle{\Omegabf_{1,N_{\rm RF}}},\dots,\angle{\Omegabf_{M_t-1,N_{\rm RF}}},\angle{\Omegabf_{M_t,N_{\rm RF}}}\right]^T$. Hence for the calibration problem, the unknown parameters can be constructed into
$\etabf = \left[\pbf^T,\theta_r,\phi_r,{\rm Real}(\gamma),{\rm Imag}(\gamma)\right]^T$.
To derive the CRB for the phase calibration problem, we have from \eqref{equ.barynx} that $\bar{\ybf} = \Dbf{\omegabf} + \bar{\zbf}$, where $\Dbf = \blkdiag\{\Cbf_1,\Cbf_2,\dots,\Cbf_{N_{\rm RF}}\}$, $\bar{\ybf} = [\bar{\ybf}_1^T,\bar{\ybf}_2^T,\dots,\bar{\ybf}_{N_{\rm RF}}^T]^T$, ${\omegabf} = [\omegabf_1^T,\omegabf_2^T,\dots,\omegabf_{N_{\rm RF}}^T]^T$, and $\bar{\zbf} = [\bar{\zbf}_1^T,\bar{\zbf}_2^T,\dots,\bar{\zbf}_{N_{\rm RF}}^T]^T$.
Hence we have
$\bar{\ybf} \sim {\mathcal CN}(\mubf, \frac{\sigma^2}{L}\Ibf)$ where $\mubf = \Dbf\omegabf$, and the Fisher Information Matrix (FIM) can be expressed as
$\Fbf_{\etabf} = \frac{2L}{\sigma^2}{\rm Re}\left\{\frac{\partial\mubf^H(\etabf)}{\partial\etabf}\frac{\partial\mubf(\etabf)}{\partial\etabf}\right\}$.
Then we have $\Cbf_{\etabf} = \Fbf_{\etabf}^{-1}$, of which the first $M_tN_{\rm RF}-1$ diagonal elements represent the corresponding CRBs of estimated variables in $\pbf$.

\subsection{Beam Pattern Optimization}\label{sec.bpo}
In this subsection, we aim for optimizing the $\fbf_{k,n}$'s to increase the accuracy of the phase estimation in \eqref{equ.objFuncGv5}. According to \eqref{equ.barynx}, the phases estimation is conducted per RF chain, thus we can focus on the $n$-th RF chain to optimize $\fbf_{k,n},k=1,2,\dots,K$. Denoting $\qbf_{n} = [\angle{\Omegabf_{1,n}},\angle{\Omegabf_{2,n}},\dots, \angle{\Omegabf_{M_t,n}}]$, we have
$\Fbf_{\qbf_{n}} =  \frac{2L}{\sigma^2}{\rm Re}\left\{\frac{\partial\xibf^H(\qbf_{n})}{\partial\qbf_{n}}\frac{\partial\xibf(\qbf_{n})}{\partial\qbf_{n}}\right\}$,
where $\xibf = \Cbf_{n}\omegabf_{n}$. We omit the constant term $\frac{2L}{\sigma^2}$ and need to solve
\begin{align}
\mathop{\min}_{\fbf_{k,n}} h({\fbf}_{k,n}) \triangleq \tr\left(\left({\rm Re}\left\{\diag(\omegabf_{n}^*)\Qbf_n\diag(\omegabf_{n})\right\}\right)^{-1}\right),
\label{eq.minT}
\end{align}
where $\Qbf_n = \beta\sum_{k=1}^K\diag(\fbf^*_{k,n})\Hbf^H\wbf_k\wbf_k^H\Hbf\diag(\fbf_{k,n})$.
Owing to the phase ambiguity removal in Section \ref{sec.pa}, we can set $\theta_t = 0$ and $\phi_t = \frac{\pi}{2}$ and thus $\abf_{\rm UPA}(\theta_t,\phi_t) = {\bf 1}_{M_t\times 1}$, where ${\bf 1}_{M_t\times 1}$ is an $M_t$-element column vector with all entries being $1$. Then
we have $\wbf_k^H\Hbf\diag(\fbf_{k,n}) = \gamma g_k\fbf_{k,n}^T$ where $g_k = \wbf_k^H\abf_{\rm UPA}(\theta_r,\phi_r)$. Thus $\Qbf_{n}$ can be simplified as
\ben
\Qbf_{n} = \beta|\gamma|^2\sum_{k=1}^K |g_k|^2 \fbf_{k,n}^* \fbf_{k,n}^T = \beta|\gamma|^2\bar{\Fbf}_{n}^*\diag(\gbf)\bar{\Fbf}_{n}^T
\een
where $\gbf = [|g_1|^2,|g_2|^2,\dots,|g_K|^2]^T$ and $\bar{\Fbf}_{n} = [\fbf_{1,n},\fbf_{2,n},\dots,\fbf_{K,n}] \in \Cnum^{M_t\times K}$. Omitting $\beta$ and $\gamma$, we denote $\Ebf_{n} =  \diag(\omegabf_{n}^*)\Qbf_n\diag(\omegabf_{n})=\diag(\omegabf_{n}^*)\bar{\Fbf}_{n}^*\diag(\gbf)\bar{\Fbf}_{n}^T\diag(\omegabf_{n})$. $\Ebf_{n}$ must be invertible so that the phase deviations are estimable, yielding that $K \ge M_t$.
Thus, the problem \eqref{eq.minT} can be expressed as
\begin{align}
\mathop{\min}_{\bar{\Fbf}_{n}} h(\bar{\Fbf}_{n}) \triangleq \tr\left(\Rbf_n^{-1}  \right).
\label{equ.minTv2}
\end{align}
where $\Rbf_n = \Ebf_{n}+\Ebf_{n}^*$.
Using the formula $\partial \Xbf^{-1} = -\Xbf^{-1}(\partial \Xbf)\Xbf^{-1}$ and $\tr(\Xbf\Ybf)=\tr(\Ybf\Xbf)$, we can obtain the Euclidean gradient as
$\nabla h(\bar{\Fbf}_{n}) = -2\diag(\omegabf_n^*)\Rbf_{n}^{-T}\Rbf_{n}^{-T}\diag(\omegabf_n)\bar{\Fbf}_{n}\diag(\gbf)$.
Similar to \eqref{equ.objFuncGv5}, we can also use the RCG algorithm to find a sub-optimal solution to \eqref{equ.minTv2} given the unit phase-shift of $\bar{\Fbf}_{n}$. In practice, the optimization of $\bar{\Fbf}_n$ cannot utilize the genie values of $\omegabf_{n}$, thus we set $\omegabf_{n} = {\bf 1}_{M_t\times 1}$ for the practical implementation of the RCG algorithm.

The SPA calibration of practical LEO communication systems can be divided into three steps: first, beam patterns are calculated during the satellite's idle periods; second, the LEO satellite sends the pilots to the ground-based terminal using these patterns; third, the terminal performs the proposed block coordinate descent algorithm, and then these estimated phases are sent back to the LEO satellite for real calibration of the SPA.

\section{Simulation Results}
This section provides some simulations to verify the performance of the OTA calibration for the SPA. In the following simulations, the complex gains $\gamma$ follow the distribution of ${\cal CN}(0,1)$; $\theta_r$, $\phi_r$, $\theta_t$, and $\phi_t$ are uniformly generated from $[-\frac{\pi}{2},\frac{\pi}{2}]$ and $[0,\pi]$, respectively; without loss of generality, the effect of pathloss $\beta$ will be included in the signal to noise ratio (SNR), and the phase deviations are uniformly distributed between $[-\varepsilon,\varepsilon]$. We define the root mean squared error $\text{RMSE} = {\sqrt{\frac{1}{M_tN_{\rm RF}-1}{\mathbb E}\{||\hat{\pbf}-{\pbf}||_2^2\}}}\times \frac{180^{\circ}}{\pi}$ as the metric to gauge the performance. Unless otherwise specified, we set $N_{\rm RF} = 4, L = 4$, $N_y = 32$, $K = 1024$, and $M_r = M_x\times M_y = 32 \times 32 = 1024$,

\begin{figure}[htb]
\centering
{\psfig{figure=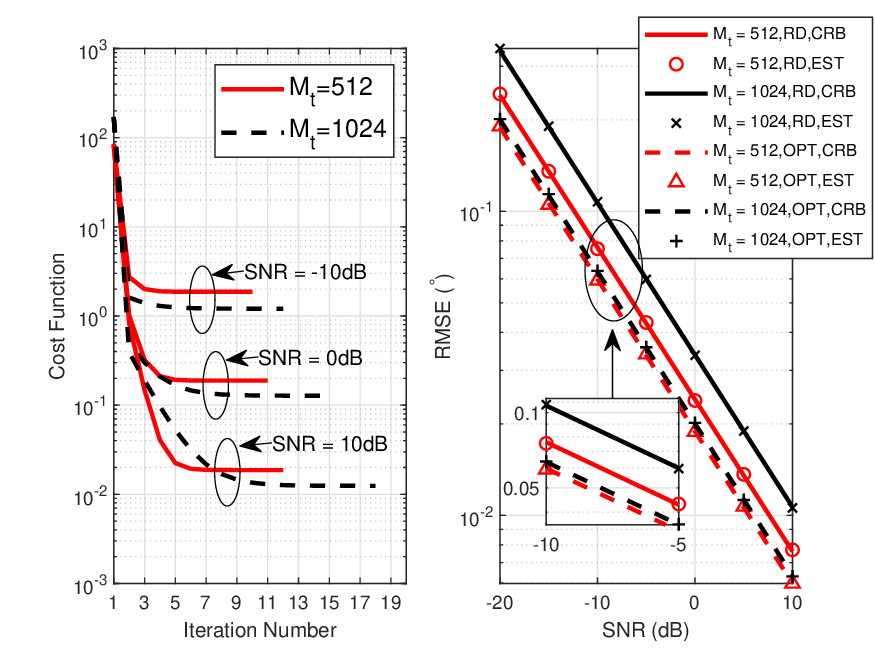,width= 3.in}}
\caption{The convergence performance and the RMSE performance under different numbers of the SPA antennas.}
\label{fig.rmseVsSNR}
\end{figure}
Fig. \ref{fig.rmseVsSNR} provides both the convergence performance and the RMSE performance of the propose algorithm given that $\varepsilon = 20^\circ$ and $N_x = 16,32$, i.e., $M_r = 512, 1024$. The left plot shows that the proposed algorithm converges in less than $10$ block descent iterations with SNR $= -10$dB, $0$dB, $10$dB. In the right plot, the RMSE performance is simulated using the random generated beam patterns (RD) and the optimized beam patterns (OPT) as SNR varies from $-20$dB to $10$dB. The right plot shows that the RMSEs closely approach the corresponding CRBs and the optimized beam patterns can offer more than $4$dB gain of SNR as $M_t = 1024$, verifying the effectiveness of both the proposed OTA calibration algorithm and the beam pattern optimization method.

\begin{figure}[htb]
\centering
{\psfig{figure=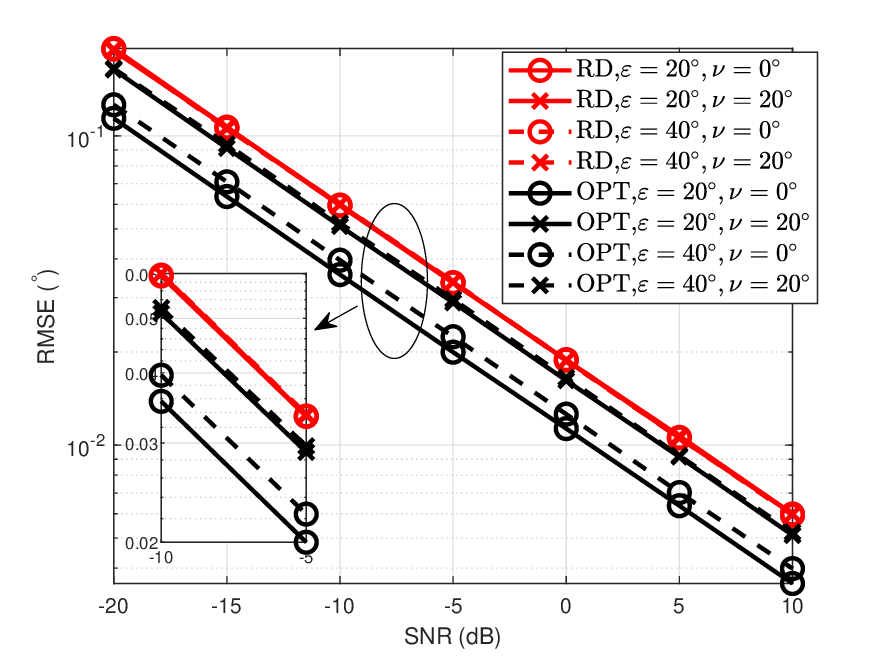,width= 3.in}}
\caption{RMSE vs. SNR under different phase deviations and DOA deviations.}
\label{fig.bp}
\end{figure}

The second example simulates the effect of the imperfect channel state information (CSI) and the phase deviations on the RMSE performance of the proposed beam pattern optimization method, where $N_x = 32$, i.e., $M_t = 1024$. For the imperfect CSI, the deviations of $\theta_r$ and $\phi_r$ are uniformly distributed between $[-\nu,\nu]$, and $\nu = 0^\circ, 20^\circ$ are considered here; for phase deviations, $\varepsilon = 20^\circ, 40^\circ$ are both considered. First, Fig. \ref{fig.bp} shows that the random beam patterns (red lines) share the same RMSE performance with $\nu = 0^\circ, 20^\circ$ and $\varepsilon = 20^\circ, 40^\circ$, indicating that the randomly generated beam patterns are the main bottleneck of the OTA phase calibration algorithm. Second, Fig. \ref{fig.bp} shows that the RMSE performance of the optimized beam patterns (black lines) varies with different deviations and generally outperforms that of the random beam patterns by at least $1$dB, where $\nu = 0^\circ, \varepsilon = 20^\circ$ has the best RMSE performance and both larger $\nu$ and $\varepsilon$ can reduce the RMSE performance, that is because $\nu = 0^\circ$ and $\varepsilon = 0^\circ$ are assumed in the practical solution to \eqref{equ.minTv2}.
\addtolength{\topmargin}{0.01in}
\section{Conclusion} \label{SEC5}
This paper considers the over-the-air (OTA) phase calibration of the spaceborne phased array (SPA) for the low earth orbit (LEO) communications. We propose to jointly calibrate the phase deviations of the SPA and estimate channel state information (CSI) using multiple pilots. The Cram\'{e}r Rao Bound (CRB) is derived and the beam patterns optimization is proposed to lower the root mean squared error (RMSE). The simulation results verify the effectiveness of the proposed algorithm as the RMSEs of the phase estimates are close to the corresponding CRBs, and the beam pattern optimization scheme is also validated for more than $4$dB gain of SNR over the randomly generated beam patterns.

\bibliographystyle{IEEEtran}
\bibliography{bib}

\end{document}